\documentclass{article}%
\usepackage{amsmath}
\usepackage{amsfonts}
\usepackage{amssymb}
\usepackage{graphicx}
\usepackage{algorithm}
\usepackage{algorithmic}%
\providecommand{\U}[1]{\protect\rule{.1in}{.1in}}
\begin{document}

\title{A simple genome-wide association study algorithm}
\author{Lev V. Utkin$^{1}$ and Irina L. Utkina$^{2}$\\$^{1}${\small Peter the Great St.Petersburg Polytechnic University, St.Petersburg, Russia}
\\{\small lev.utkin@gmail.com, utkin\_lv@spbstu.ru}
\\$^{2}${\small Skolkovo Institute of Science and Technology, Skolkovo, Moscow Region, Russia}
\\$^{2}${\small Peter the Great St.Petersburg Polytechnic University, St.Petersburg, Russia}
\\{\small ira.l.utkina@gmail.com, irina.utkina@skolkovotech.ru}
}
\date{}

\maketitle

\begin{abstract}
A computationally simple genome-wide association study (GWAS)
algorithm for estimating the main and epistatic effects of markers or single
nucleotide polymorphisms (SNPs) is proposed. It is based on the intuitive
assumption that changes of alleles corresponding to important SNPs in a pair
of individuals lead to large difference of phenotype values of these
individuals. The algorithm is based on considering pairs of individuals
instead of SNPs or pairs of SNPs. The main advantage of the algorithm is that
it weakly depends on the number of SNPs in a genotype matrix. It mainly
depends on the number of individuals, which is typically very small in
comparison with the number of SNPs. Numerical experiments with real data sets
illustrate the proposed algorithm.


\end{abstract}

\section{Introduction}

A genome-wide association study (GWAS) aims to discover genetic factors
underlying phenotypic traits, i.e., GWAS examines the association between
phenotypes and genetic variants or genotypes across the entire genome. It can
be regarded as one of the methods for the well-known feature selection problem
where features are the so-called single nucleotide polymorphisms (SNPs). SNPs
are typically used as markers of a genomic region and can be defined as a DNA
sequence variation where a single nucleotide (A, T, C, G) in the genomic
sequence differs among the individuals of a biological species. It should be
noted that most SNPs have no effect on the phenotype values or their effect is
very insignificant. However, there are SNPs which might be very important in
associations between SNPs and the phenotypes. Therefore, another formulation
of the main aim of GWAS is to identify or select the most relevant SNPs which
differentiate one group of individuals from another or which contribute to the
phenotypic differences among the individuals.

We point out some difficulties of solving the GWAS problem mentioned by many
authors. First of all, the number of SNPs $p$ is usually very large. It is
typically 10--100 times the number of individuals $n$ in the training sample.
This is the so called $p>n$ (or large $p$ small $n$) problem. Second, genetic
mechanisms might involve complex interactions among genes and between genes
and environmental conditions which are not fully captured by additive models
\cite{Campos-Gianola-Allison-10,Goddard-09}. SNPs may interact in their
effects on phenotype, i.e., there is the so-called epistatic effect. Third,
many genetic variants are not genotyped, i.e., there are missing data in the
genotype information. Fourth, GWAS is applied to find the association between
SNPs and different kinds of the trait. It is mentioned by Zhang et al.
\cite{Zhang-Huang-Zhang-Wang-2012} in their interesting review of the GWAS
methods that the successful GWAS methods applied to identifying SNPs
contributing a disease (the two-valued or case-control phenotype) may have
problems in finding SNPs associated with complex traits (quantitative or
continuous phenotype).

A huge amount of the statistical procedures and methods solving the GWAS
problem have been developed the last decades. A part of methods can be
referred to as filter methods \cite{Altidor-2011} which use statistical
properties of SNPs to filter out poorly informative ones. The Fisher
criterion, Pearson $\chi^{2}$-test, Cochran-Armitage test are the well-known
statistical methods for detecting differential SNPs between two samples. These
methods can be joined as the so-called single-locus association tests because
the tests are performed separately for each SNP when the case-control
phenotypes is analyzed. For quantitative phenotypes, a standard tool is the
one-way ANOVA \cite{Wu-Ma-Gasella-2007}. Another part of methods uses various
kinds of regression models which can be referred to as embedded methods. One
of the pioneering papers devoted to the use of regression models in SNP
selection has been written by Lander and Botstein \cite{Lander-Botstein-89}.
The regression models mainly include the Ridge regression and Lasso
techniques, their combination called the elastic nets
\cite{Hastie-Tibshirani-Friedman09}. Comprehensive reviews of the methods and
algorithms using the regression models and their various modifications for
solving the GWAS problems are provided by Ayers and Cordell
\cite{Ayers-Cordell-10}, by Hayes \cite{Hayes-2013}.

It has been mentioned that the standard GWAS analyzes each SNP separately in
order to identify a set of significant SNPs showing genetic variations
associated with the trait. However, an important challenge in the analysis of
genome-wide data sets is taking into account the so-called epistatic effect
when different epistatic loci interact in their association with phenotype.
The epistatic effect can be viewed as gene-gene interaction when the action of
one locus depends on the genotype of another locus. At the same time, there
are different interpretations of the epistatic effect. A fundamental critical
review of different definitions and interpretations of epistasis is provided
by Cordell \cite{Cordell-2002}. From the statistical point of view, the
epistatic effect is the statistical deviation from the joined effects of two
loci on the phenotype \cite{Wan-Yang-Yang-Xue-Fan-Tang-Yu-10}. There is a
series of interesting methods which use the statistical tests at their first
step in order to reduce the set of SNPs. These are FastANOVA
\cite{Zhang-Zou-Wang-2008}, FastChi \cite{Zhang-Zou-Wang-2009}, COE
\cite{Zhang-Pan-Xie-Zou-Wang-2009}, TEAM \cite{Zhang-Huang-Zou-Wang-2010}. We
can also point out methods which differs from the filter methods, for example,
the Bayesian epistasis association mapping method (BEAM) proposed by Zhang and
Liu \cite{Zhang-Liu-2007}, tree-based methods like the random forests
\cite{Li-Horstman-Chen-2011}, the multifactor dimensionality reduction
\cite{Ritchie-2001}, modifications of the Lasso techniques
\cite{Bocianowski-2014}. Comparative analyses of methods devoted to the
epistatic interaction effect were provided by several authors
\cite{Chen-2011,Wang-Liu-Feng-Wong-2011}. Analyzing these methods, we have to
conclude that most of them have two steps (except for the methods with
exhaustive consideration of all SNP pairs) such that the first step is for
reducing the set of all SNPs to the most important ones, and the second step
solves the SNP-SNP interaction problem.

From many approaches for solving the GWAS taking into account the epistatic
effect, we would like to mark out a very interesting and efficient algorithm
\cite{Achlioptas-2011} that is subquadratic in the number of SNPs $p$. The
authors \cite{Achlioptas-2011} propose an algorithm for efficiently retrieving
some predefined number of top scoring pairs among all pairs of SNPs, assuming
binary phenotypes and the difference-in-correlation as the association
criterion. Some implicit ideas of the algorithm will be used below.

In the present study, we propose a computationally extremely simple GWAS
algorithm. It is based on the intuitive assumption that changes of alleles
corresponding to important SNPs in a pair of individuals lead to large
difference of phenotype values of these individuals. The main advantage of the
algorithm is that it weakly depends on the number of SNPs in a genotype
matrix. It mainly depends on the number of individuals, which is typically
very small in comparison with the number of SNPs. We called the algorithm FAPI
(Fast Analysis of Pairs of Individuals).

\section{The proposed algorithm}

We start with the following general definition of the association mapping
problem. Let $\mathbf{X}=[X_{1},...,X_{p}]$ be a genotype matrix for $n$
individuals and $p$ SNPs. From a statistical point of view, the genotype
matrix can be treated as a predictor matrix and the marker genotypes as
qualitative explanatory variables, i.e., $X_{j}=(x_{1j},...,x_{nj}%
)^{\mathrm{T}}$ is a predictor representing the $j$-th SNP, $j=1,...,p$. For
bi-allelic SNPs, every $x_{ij}$ is an allele of the $i$-th individual at the
$j$-th SNP locus. It can be represented by symbols $\{0,1\}$, where $0$ and
$1$ stand for majority and minority alleles, respectively. A genotype may also
be represented with any of the numbers $\{0,1,2\}$ to represent the homozygous
major allele (\textquotedblleft$AA=0$\textquotedblright), heterozygous allele
(\textquotedblleft$Aa/aA=1$\textquotedblright), and homozygous minor allele
(\textquotedblleft$aa=2$\textquotedblright), respectively. A vector of alleles
corresponding to the $i$-th individual will be denoted as $\mathbf{x}%
_{i}^{\mathrm{T}}=(x_{i1},...,x_{ip})$, $i=1,...,n$. A quantitative trait of
interest or a set of the phenotype values $y_{i}\in\mathbb{R}$, $i=1,...,n$,
can be regarded as the response vector $Y=(y_{1},...,y_{n})^{\mathrm{T}}$. The
goal of GWAS is to find SNPs in $\mathbf{X}$, that are highly associated with
$Y$, which will be called as important or significant SNPs.

The main idea underlying the FAPI is based on comparison of genotypes of pairs
of individuals and comparison of the corresponding phenotype values. At that,
we use the following intuitive assumption. If genotypes of two individuals are
close to each other and the corresponding phenotype values of these two
individuals are far from each other, then the SNP-markers which correspond to
different elements of the considered two genotypes \textit{might be} important
or contribute to the phenotype values. Indeed, if two individuals differ by
some small number of genotype elements, then it is naturally to expect that
their phenotypes are similar. However, if the corresponding phenotypes are
substantially different, then it is naturally to suppose that this\ small
number of \textquotedblleft distinguishing\textquotedblright\ genotype
elements define this large difference of phenotypes values. Of course, the
large difference of the phenotype values may be caused by the noise or other
random factors. Therefore, we cannot make any conclusions only on the basis of
one pair of individuals. That is why the word combination \textquotedblleft%
\textit{might be}\textquotedblright\ used above\ means that this assumption
may be wrong due to random character of the phenotype values. But we can make
the conclusion by analyzing all pairs of individuals or a part of all pairs.

Informally, the FAPI can be written as follows. First of all, we find all
pairs $(\mathbf{x}_{i},\mathbf{x}_{j})$ of vectors of alleles. Then, we select
some predefined number of the pairs which have largest differences of
phenotype values and smallest distances between the vectors of alleles for
every pair in accordance with some combined measure jointly characterizing the
differences and the distances. The next step is to make a decision which SNPs
contribute to the difference between the vectors of alleles for the
\textquotedblleft best\textquotedblright\ pairs. The use of the predefined
number of pairs allows us to smooth possible outliers of the phenotype values
due to random factors.%

\begin{table}[tbp] \centering
\caption{Genotypes and phenotypes of three individuals}%
\begin{tabular}
[c]{ccccc}\hline
\multicolumn{2}{c}{Phenotypes} & 45 & 15 & 10\\\hline
& 1 & 0 & 0 & 1\\\cline{2-5}
& 2 & 0 & 0 & 1\\\cline{2-5}%
SNPs & 3 & 1 & 0 & 0\\\cline{2-5}
& 4 & 1 & 1 & 0\\\cline{2-5}
& 5 & 1 & 1 & 1\\\hline
\end{tabular}
\label{t:SNP_dist_1}%
\end{table}%
%

\begin{table}[tbp] \centering
\caption{The genotype transitions and the phenotype differences of three
pairs, values of $\rho $ and $r$}%
\begin{tabular}
[c]{ccccc}\hline
\multicolumn{2}{c}{$d(y_{i},y_{j})$} & 30 & 35 & 5\\\hline
& 1 & 0$\rightarrow$0 & 0$\rightarrow$1 & 0$\rightarrow$1\\\cline{2-5}
& 2 & 0$\rightarrow$0 & 0$\rightarrow$1 & 0$\rightarrow$1\\\cline{2-5}%
SNPs & 3 & 1$\rightarrow$0 & 1$\rightarrow$0 & 0$\rightarrow$0\\\cline{2-5}
& 4 & 1$\rightarrow$1 & 1$\rightarrow$0 & 1$\rightarrow$0\\\cline{2-5}
& 5 & 1$\rightarrow$1 & 1$\rightarrow$1 & 1$\rightarrow$1\\\hline
$\rho(\mathbf{x}_{i},\mathbf{x}_{j})$ &  & 1 & 4 & 3\\\hline
$r(i,j)$ &  & 30 & 8.75 & 1.667\\\hline
\end{tabular}
\label{t:SNP_dist_2}%
\end{table}%
%

\begin{table}[tbp] \centering
\caption{Vectors $\bf{z}_{ij}$ and decision making about the important SNP
(the third SNP)}%
\begin{tabular}
[c]{cccc}\hline
\multicolumn{2}{c}{$r^{\ast}(i,j)$} & 30 & 8.75\\\hline
& 1 & 0 & -1\\\cline{2-4}
& 2 & 0 & -1\\\cline{2-4}%
SNPs & \multicolumn{1}{||c}{\textbf{3}} & \multicolumn{1}{||c}{\textbf{1}} &
\textbf{1}\\\cline{2-4}
& 4 & 0 & 1\\\cline{2-4}
& 5 & 0 & 1\\\hline
\end{tabular}
\label{t:SNP_dist_4}%
\end{table}%

Formally, the proposed algorithm FAPI can be represented as follows.

\begin{enumerate}
\item All vectors of alleles $\mathbf{x}_{1},...,\mathbf{x}_{n}$ are sorted in
descending order of the corresponding phenotypes, i.e., $y_{1}\geq...\geq
y_{n}$. This step is not necessary, but it simplifies comparison of phenotype
values, namely, the condition $y_{i}-y_{j}\geq0$ for all $i<j$ is valid in
this case.

\item All different pairs of individuals are composed. The number of pairs is
$n(n-1)/2$. Only pairs $(\mathbf{x}_{i},\mathbf{x}_{j})$ such that $i<j$ are considered.

\item For every pair $\mathbf{x}_{i}$, $\mathbf{x}_{j}$, the distance
$\rho(\mathbf{x}_{i},\mathbf{x}_{j})$ between vectors $\mathbf{x}_{i}$ and
$\mathbf{x}_{j}$, $i,j=1,...,n$, $i<j$, is computed. A type of the distance
depends on data. It can be the standard Hamming distance for binary variables
$x_{ij}\in\{0,1\}$. The standard Euclidean distance metric can be also used here.

\item For every pair $(i,j)$, the difference $d(y_{i},y_{j})$ between
phenotype values $y_{i}$ and $y_{j}$, $i,j=1,...,n$, $i<j$, is computed. The
condition $d(y_{i},y_{j})\geq0$ is valid because phenotypes are sorted in
descending order (see Step 1).

\item For every pair $(i,j)$, the ratio
\[
r(i,j)=d(y_{i},y_{j})/\rho(\mathbf{x}_{i},\mathbf{x}_{j})
\]
is computed. The larger the difference $d$ and the smaller the distance $\rho$
are, the larger ratio $r$ is. The ratio $r$ is a measure of target pairs.

\item $N$ largest values of $r(i,j)$ are selected. Denote these values as
$r^{\ast}(i,j)$ and the set of their indices $(i,j)$ as $J^{\ast}$. The value
$N$ can be regarded as a tuned parameter later. Another way is to compute the
value $N$ by constructing a cumulative probability distribution of the random
variable $R$ whose sample values are $r(i,j)$. It was observed by many
numerical experiments that values $r(i,j)$ have a unimodal distribution.
Moreover, if we assume that random variables taking values $d(y_{i},y_{j})$
and $\rho(\mathbf{x}_{i},\mathbf{x}_{j})$ have some distributions, for
example, normal distributions, then $R$ has one of the so-called ratio
distributions, for example, the Cauchy distribution, the $t$-distribution, the
$F$-distribution. Therefore, we take a predefined value of $q\%$ quantile of
the random variable $R$ and find all values of the ratio such that their
empirical distribution function is larger than $q/100$. In this case, we
derive some value of $N$ from the above procedure, and $q$ can be viewed as a
tuned parameter of the algorithm.

\item For every pair $(i,j)$ from $J^{\ast}$, we find a subset of elements of
vectors $\mathbf{x}_{i}$ and $\mathbf{x}_{j}$ which differentiate these
vectors. In particular, if $x_{ij}\in\{0,1\}$, then we find the vector
$\mathbf{z}_{ij}=\mathbf{x}_{i}-\mathbf{x}_{j}$. The vector $\mathbf{z}_{ij}$
has element $-1$ at the $k$-th position if there is the transition from $0$ in
$\mathbf{x}_{i}$ to $1$ in $\mathbf{x}_{j}$ at the $k$-th position, element
$1$ if there is the transition from $1$ in $\mathbf{x}_{i}$ to $0$ in
$\mathbf{x}_{j}$ at the same position, and element $0$ by transitions from $0$
to $0$ or from $1$ to $1$ at the same position, i.e.,%
\[
\mathbf{z}_{ij}(k)=\left\{
\begin{array}
[c]{cc}%
-1, & \text{if }\mathbf{x}_{i}(k)=0,\ \mathbf{x}_{j}(k)=1,\\
1, & \text{if }\mathbf{x}_{i}(k)=1,\ \mathbf{x}_{j}(k)=0,\\
0, & \text{if }\mathbf{x}_{i}(k)=\mathbf{x}_{j}(k).
\end{array}
\right.
\]

Only elements of $\mathbf{z}_{ij}$ with values $-1$ and $1$ are interesting
for us because they indicate positions where vectors $\mathbf{x}_{i}$ and
$\mathbf{x}_{j}$ are different, which, in turn, indicate
the\textit{\ possible} important SNPs. In the case $x_{ij}\in\{0,1,2\}$, we
have six transitions $0\rightarrow1$, $0\rightarrow2$, $1\rightarrow0$,
$1\rightarrow2$, $2\rightarrow0$, $2\rightarrow1$ enumerated as
$-3,-2,-1,1,2,3$, and three transitions $0\rightarrow0$, $1\rightarrow1$,
$2\rightarrow2$ denoted as $0$.

\item For the $k$-th SNP under condition $x_{ij}\in\{0,1\}$, we use the ratio
$r^{\ast}(i,j)$ for computing summed weights of elements $-1$, $0$, $1 $ at
the $k$-th position in $\mathbf{z}(i,j)$ denoted as $a_{k}(-1)$, $a_{k}(0)$,
$a_{k}(1)$, i.e., we compute
\[
a_{k}(t)=\sum_{(i,j)\in J^{\ast}}r_{\text{norm}}^{\ast}(i,j)\mathbf{1}%
(\mathbf{z}_{ij}(k)=t),\ t=-1,0,1.
\]
Here $\mathbf{1}(\mathbf{z}_{ij}(k)=t)$ is the indicator function taking the
value $1$ if $\mathbf{z}_{ij}(k)=t$, and the value $0$ otherwise;
$r_{\text{norm}}^{\ast}$ is the normalized ratio. We can also take
$r_{\text{norm}}^{\ast}(i,j)=1$ for all values $(i,j)\in J^{\ast}$. In this
simplified case, we just find the numbers of elements $-1$, $0$, $1$ at the
$k$-th position in vectors $\mathbf{z}_{ij}$, $(i,j)\in J^{\ast}$. If
$x_{ij}\in\{0,1,2\}$, then $t$ takes values from set $T=\{-3,-2,-1,0,1,2,3\}$.

\item For the $k$-th SNP, we compare two numbers $a_{k}(t=0)$ and $\sum
_{t\neq0}a_{k}(t)$ with a decision threshold $h$, i.e., we compare the summed
weights of transitions which do not contribute to the difference of vectors of
alleles and which correspond to transitions $0\rightarrow0$, $1\rightarrow1$,
and the weights of transitions with different values of alleles. If the
inequality
\[
\sum_{t\neq0}a_{k}(t)/a_{k}(t=0)\geq h
\]
is valid, then the corresponding $k$-th SNP is important, otherwise it does
not belong to the subset of important SNPs. The decision threshold is
typically equal to $1$.

\item For every target SNP, we write the value
\[
S_{k}=\arg\max_{t\in\{-1,1\}}a_{k}(t).
\]
Values $S_{k}=-1$ or $1$ mean that the allele corresponding to the $k$-th SNP
and having values $1$ or $0$, respectively, contributes to decreasing of the phenotype.
\end{enumerate}

Let us illustrate the above algorithm by means of a toy example. Suppose we
have $n=3$ individuals whose genotype matrix for $5$ bi-allelic SNPs is
represented by symbols $0$ and $1$ which stand for major and minor alleles,
respectively. The sorted phenotype values are $45$, $15$, $10$. The initial
data are shown in Table \ref{t:SNP_dist_1}. We have three pairs of vectors of
alleles such that the phenotype differences $d(y_{i},y_{j})$, the genotype
transitions, the corresponding Hamming distances between vectors of alleles in
every pair and the ratios $r(i,j)$ are given in Table \ref{t:SNP_dist_2}.
Suppose that the threshold $N$ for selecting the largest values of $r(i,j)$ is
$2$. Table \ref{t:SNP_dist_4} shows individuals satisfying this condition and
the values $\mathbf{z}(i,j)$ of transitions taking the values $-1,0,1$ (see
Step 7). It can be seen from Table \ref{t:SNP_dist_4} that only the third SNP
has two non-zero elements $\mathbf{z}(i,j)$. This implies that only the third
SNP is important. Indeed, it is obviously from Table \ref{t:SNP_dist_1} that
the largest difference is observed between phenotypes of the first and the
second individuals. Moreover, only the third SNP separates the first and the
second vectors of alleles. Intuitively, we can conclude that this SNP is a
reason for the large difference between phenotypes of the first and the second individuals.

The FAPI for determining important SNPs is represented as Algorithm
\ref{alg:FAPI}.

\begin{algorithm}
\caption{A simple FAPI algorithm} \label{alg:FAPI}
\begin{algorithmic}
[1]\REQUIRE$\mathbf{X}_{n\times p}=(\mathbf{x}_{1},...,\mathbf{x}_{n})$
(binary genotype matrix), $Y$ (phenotype vector), $N$, $h$ (parameters)
\ENSURE$S_{k} $ (imported SNPs) \STATE Order $(\mathbf{x}_{1},...,\mathbf{x}%
_{n})$ such that $y_{1}\geq...\geq y_{n}$
\FOR{each $i\leq n$, $j>i$ } \STATE Build a pair $(\mathbf{x}_{i}%
,\mathbf{x}_{j})$ \STATE Compute $r(i,j)=(y_{i}-y_{j})/\rho(\mathbf{x}%
_{i},\mathbf{x}_{j})$ \STATE Compute $\mathbf{z}_{ij}=\mathbf{x}%
_{i}-\mathbf{x}_{j}$ \ENDFOR
\STATE$J^{\ast}=\{(i,j):N$ largest values of $r(i,j)\}$
\FOR{each $k\leq p$ } \STATE Compute $a_{k}(t)=\sum_{(i,j)\in J^{\ast}%
}r_{\text{norm}}^{\ast}(i,j)\mathbf{1}(\mathbf{z}_{ij}(k)=t),\ t=-1,0,1.$
\IF{$\sum_{t\neq 0}a_{k}(t)\geq h\cdot a_{k}(t=0)$} \STATE$S_{k}$ is important
\ENDIF
\ENDFOR
\end{algorithmic}
\end{algorithm}

\section{Properties of the algorithm}

Let us point out some properties and advantages of the FAPI.

\begin{enumerate}
\item The epistatic effect which is viewed as gene-gene interaction should not
be separately analyzed. It is implicitly included into the proposed algorithm.
Indeed, we do not consider single SNPs. For every pair of vectors of alleles,
the difference of the vectors is computed for all SNPs simultaneously. So, if
there is a combination of alleles which significantly impact on the phenotype,
it produces a large difference between the corresponding phenotype values.
This is a very important property which allows us to significantly reduce the
computational burden needed for consideration of many SNP pairs.

\item The FAPI is very simple. Its computational complexity is $O(p+n^{2})$,
i.e. the complexity is linear with the number of SNPs $p$. This is a very
important property of the algorithm because the number of SNPs $p$ is
typically 10--100 times the number of individuals $n$ in the training sample
for many problems. Moreover, the algorithm does not require special procedures
like Lasso, etc.

\item The FAPI does not depend on the set of allele values. For example, a few
trivial changes are needed to consider the case $x_{ij}\in\{0,1,2\}$.
Moreover, the important feature of the algorithm is that the values
$\{0,1,2\}$ or $\{0,1\}$ are viewed as categorical numbers without order, for
example, $0<1<2$. The FAPI can be modified for the case $x_{ij}\in\mathbb{R} $
which takes place in the microarray gene expression data analysis.

\item Another advantage of the FAPI is handling missing data in the genotype
matrix. We do not need to apply special procedures for pre-processing missing
data and their imputation. The missing data just extend the set of values of
every $x_{ij}$. We use the conservative strategy. For example, suppose
$x_{ij}\in\{0,1\}$ and the missing value is denoted as $2$. If we have two
missing values at the same $k$-th position in vectors $\mathbf{x}_{i}$ and
$\mathbf{x}_{j}$, then $\mathbf{z}_{ij}(k)=0$. This value means that we do not
consider the $k$-th position in vectors $\mathbf{x}_{i}$ and $\mathbf{x}_{j}$
as a candidate for getting an important SNP. At the same time, when we have a
single missing value at the $k$-th position in vectors $\mathbf{x}_{i}$ and
$\mathbf{x}_{j}$, then $\mathbf{z}_{ij}(k)\neq0$ in accordance with the
strategy that a larger number of important SNPs is preferable because the
second selection from a small subset of important SNPs should be carried out
by means of the well-known standard procedures.

\item The FAPI can be used when the phenotype takes only two values (the
case-control study). It is obvious in this case that only a set composed from
pairs of individuals taken from the case and control groups, respectively, is
analyzed. Indeed, $d(y_{i},y_{j})=0$, $r(i,j)=0$ if $y_{i}=y_{j}$, and
$d(y_{i},y_{j})=1$, $r(i,j)=1/\rho(\mathbf{x}_{i},\mathbf{x}_{j})$ if
$y_{i}\neq y_{j}$ (we assume that the vectors of alleles are sorted in
descending order of the corresponding phenotypes).

\item For many available algorithms of GWAS using filter methods for selection
of the most important SNPs like the Fisher exact test, the one-way ANOVA, etc.
we have to predefine a limit number of the important SNPs. The FAPI determines
this number itself.

\item The FAPI can be tuned by means of the parameter $N$ (the number of
largest values of the rate $r$) or parameter $q$. On the one hand, too small
values of the parameter $N$ may lead to a large number of target SNPs. As a
result, we have to use some additional procedures for restricting the number
of SNPs. On the other hand, large values of $N$ may lead to possible missing
SNPs which actually may be very important. There is a compromise choice of $N
$ which can be carried out by considering all possible values of $N$ in a
predefined grid. Another parameter for tuning is the decision threshold $h$.

\item The FAPI is flexible. This means that many its elements can be changed.
For example, there are many metrics for computing distances between vectors of
alleles such that the choice of an appropriate metric might improve the
algorithm. Similarity $S(\mathbf{x}_{i},\mathbf{x}_{j})$ and dissimilarity
$\rho(\mathbf{x}_{i},\mathbf{x}_{j})$ measures of two vectors $\mathbf{x}_{i}$
and $\mathbf{x}_{j}$ can be applied. If we use similarity measures, then
$r(i,j)=d(y_{i},y_{j})\cdot S(\mathbf{x}_{i},\mathbf{x}_{j})$. Another element
which could be changed is the choice of the ratio $r$. The proposed ratio is
one of the possible measures for the target pairs localization. It is just the
most simple way for defining the measure. Perhaps, other measures might also
improve the algorithm.
\end{enumerate}

\section{Numerical experiments}

\subsection{Data sets}

Numerical experiments are carried out on three populations of double haploid
(DH) lines of barley:

\begin{enumerate}
\item The first dataset consists of 175 DH lines of barley
\cite{Chutimanitsakun-2011,Cistue-2011}. The correponding phenotyping and
genotyping data are available at Oregon Wolfe Barley Data (OWBD) and
GrainGenes Tools \newline(http://wheat.pw.usda.gov/ggpages/maps/OWB/).
\newline The lines are analyzed with respect to the heading date trait. The
linkage map consists of 1328 SNPs.

\item The second dataset consists of 92 DH lines of barley from the Dicktoo x
Morex cross and described by Hayes et al. \cite{Hayes-1993a,Hayes-1997}, by
Pan et al. \cite{Pan-1994}. The corresponding data are available at \newline
http://wheat.pw.usda.gov/ggpages/DxM/ . \newline We analyze the lines with
respect to two phenotypic traits: heading date with and without vernalization
with an 8-h light/16-h dark photoperiod regime. The linkage map consists of
117 SNPs.

\item The third population dataset includes 150 DH lines of barley from the
Steptoe x Morex cross \cite{Close-2009,Hayes-Jyambo-1993}. The corresponing
data are available at \newline http://wheat.pw.usda.gov/ggpages/SxM. \newline
The linkage map consists of 223 SNPs. The lines are analyzed with respect to
the heading date trait measured in 16 environments and grain yield trait
measured in 6 environments.
\end{enumerate}

The missing data are handled by means of extending the set of values of every
$x_{ij}$, i.e., the set of values $\{0,1\}$ is extended on the set $\{0,1,2\}$.

\subsection{The first dataset}

\begin{figure}
[ptb]
\begin{center}
\includegraphics[
height=2.2833in,
width=4.5236in
]%
{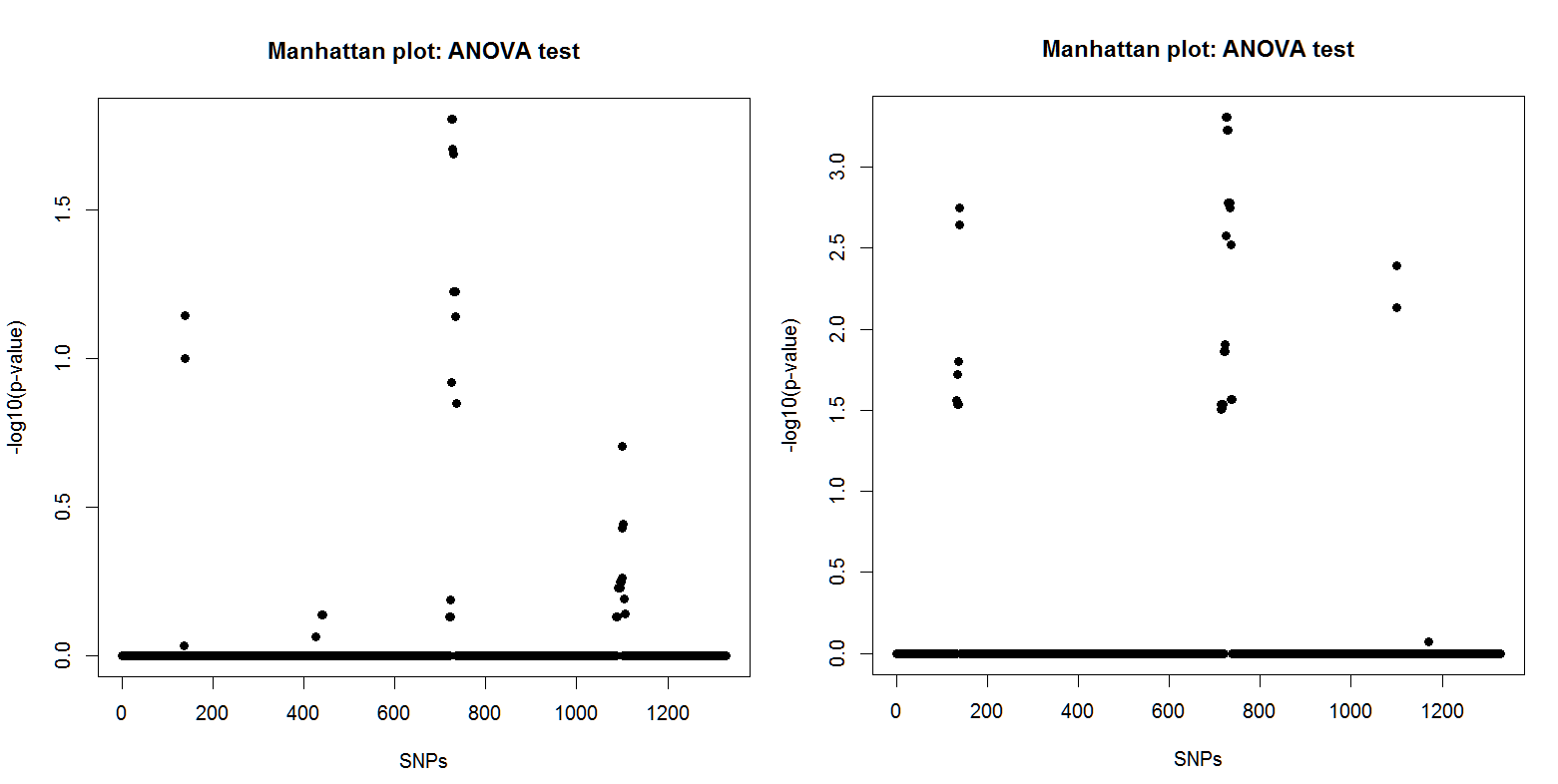}%
\caption{The Manhattan plot for the OWBD using standard method (left) and the
FAPI (right)}%
\label{fig:Manh_plot_12}%
\end{center}
\end{figure}

First, we investigate DH lines of barley from OWBD. The parameter $q$ is
$97\%$. In order to compare the proposed algorithm, we apply the standard tool
ANOVA to testing the association between a single marker and a continuous
outcome. The F-test is used to assess whether the expected values of a
quantitative variable within several predefined groups differ from each other.
From this, we can retrieve a p-value for the significance of association
between each SNP and the phenotype. Then we correct for multiple testing using
the Holm--Bonferroni method. The Manhattan plot generated from the obtained
p-values is shown in Fig. \ref{fig:Manh_plot_12} (the left plot). One can see
from the Fig. \ref{fig:Manh_plot_12} that the significant SNPs have numbers
close to 139, 725, 1100. SNPs with these numbers have the smallest p-values.

Let us look at Fig. \ref{fig:Manh_plot_12} (the right plot) now. It shows a
similar Manhattan plot, but significant SNPs are obtained by using the FAPI,
and p-values are computed for this set again using the Holm--Bonferroni
correction. However, the first step of the FAPI provides not only the
significant SNPs which coincide with the SNPs derived by the standard tool
ANOVA. It provides SNPs with numbers 1169 and 1302, which do not belong to the
set of significant SNPs obtained by means of the ANOVA. It turns out that the
p-values of these single SNPs are larger than $0.05$, i.e., they cannot be
viewed as significant ones. In contrast to the single-locus approach applied
before, we perform the ANOVA test in order to identify interacting SNP-pairs
that have strong association with the phenotype. It is important to note that
the two-locus ANOVA test is performed on a small number of candidate SNP-pairs
which have been obtained by means of the FAPI. It turns out that SNPs with
numbers 1169 and 1302 interact with SNPs 729 and 725, respectively, such that
the corresponding p-values ($0.021$ and $0.047$) after the Holm--Bonferroni
correction are smaller than $0.05$. In other words, the FAPI\ allows us to
implement the efficient epistasis detection.

\subsection{The second dataset}

\begin{figure}
[ptb]
\begin{center}
\includegraphics[
height=2.3176in,
width=4.5508in
]%
{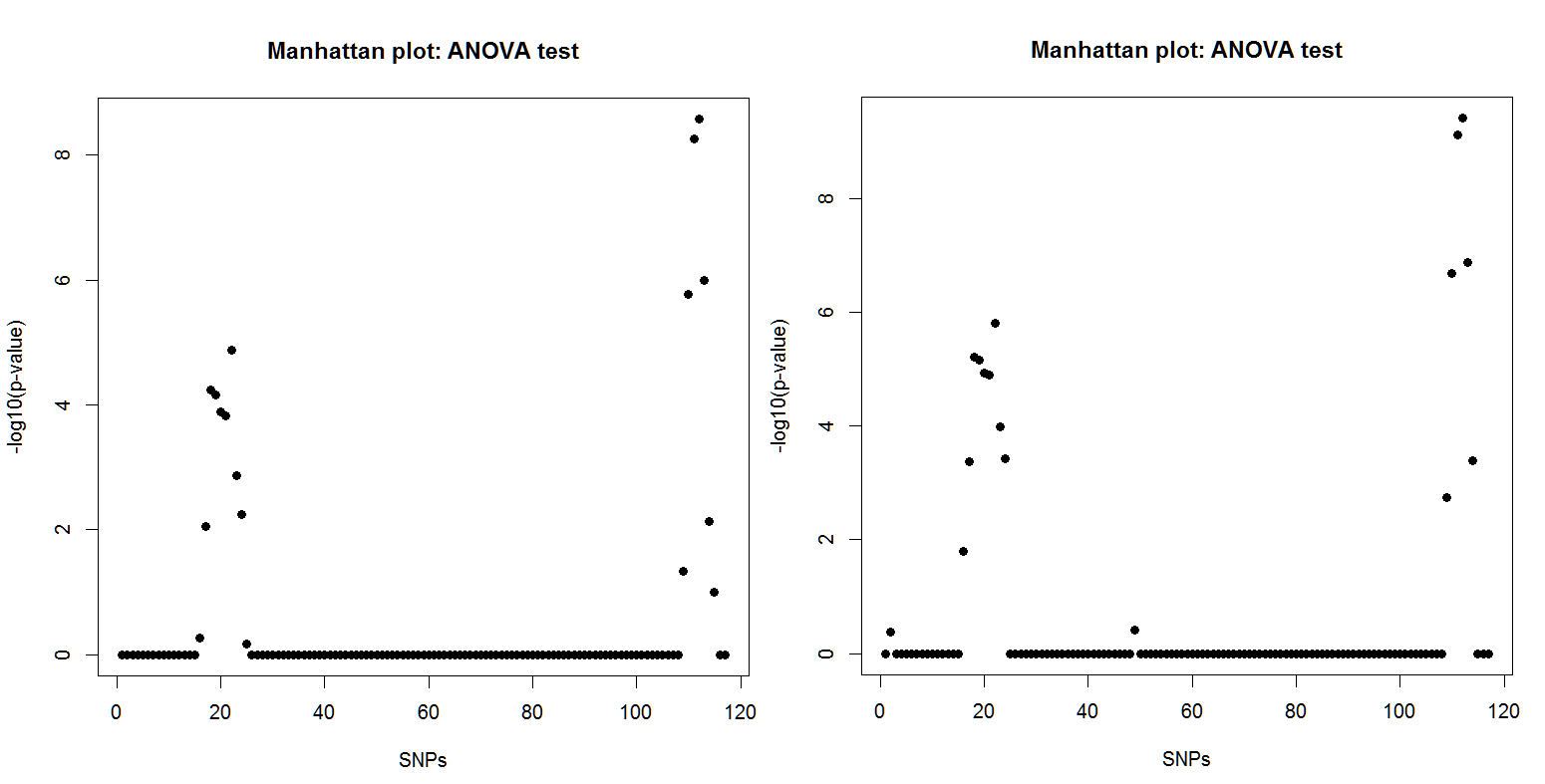}%
\caption{The Manhattan plots for the Dicktoo x Morex data set using standard
method (left) and the FAPI (right)}%
\label{fig:Manh_plot_34}%
\end{center}
\end{figure}

Let us study the dataset obtained from the Dicktoo x Morex cross. According to
Pan et al. \cite{Pan-1994} (Page 905), top ranked SNPs for heading date with
and without vernalization are ABC170-CD064 and Dhn1-BCD265b which correspond
to the following numbers of SNPs 22-24 and 111-113, respectively. The ANOVA is
applied here again. We get two SNPs with numbers 22 and 112 having the
smallest p-values $1.32\times10^{-5}$ and $2.66\times10^{-9}$, respectively.
The corresponding Manhattan plot is shown in Fig. \ref{fig:Manh_plot_34} (the
left plot). Numerical experiments with using the FAPI provide quite the same
results. They are shown in Fig. \ref{fig:Manh_plot_34} (the right plot).
However, the FAPI indicates that there is the 49-th SNP (saflp35) which has a
large p-values, but its interaction with SNPs 112 and 22 gives the p-values
0.0135 and 0.0144, respectively. All p-values are computed by using the
Holm--Bonferroni correction.

We get similar results for the unvernalized treatment (the second phenotypic
trait). In addition, we obtain SNPs with numbers 36, 59, 76, which are called
as saflp219, SOLPRO, HorB, respectively, and which are located on different
chromosomes. These SNPs interact with the SNP 22 with the corresponding
p-values $0.0034$, $0.038$, $0.045$, respectively.

\subsection{The third dataset}

\begin{figure}
[ptb]
\begin{center}
\includegraphics[
height=2.3299in,
width=4.571in
]%
{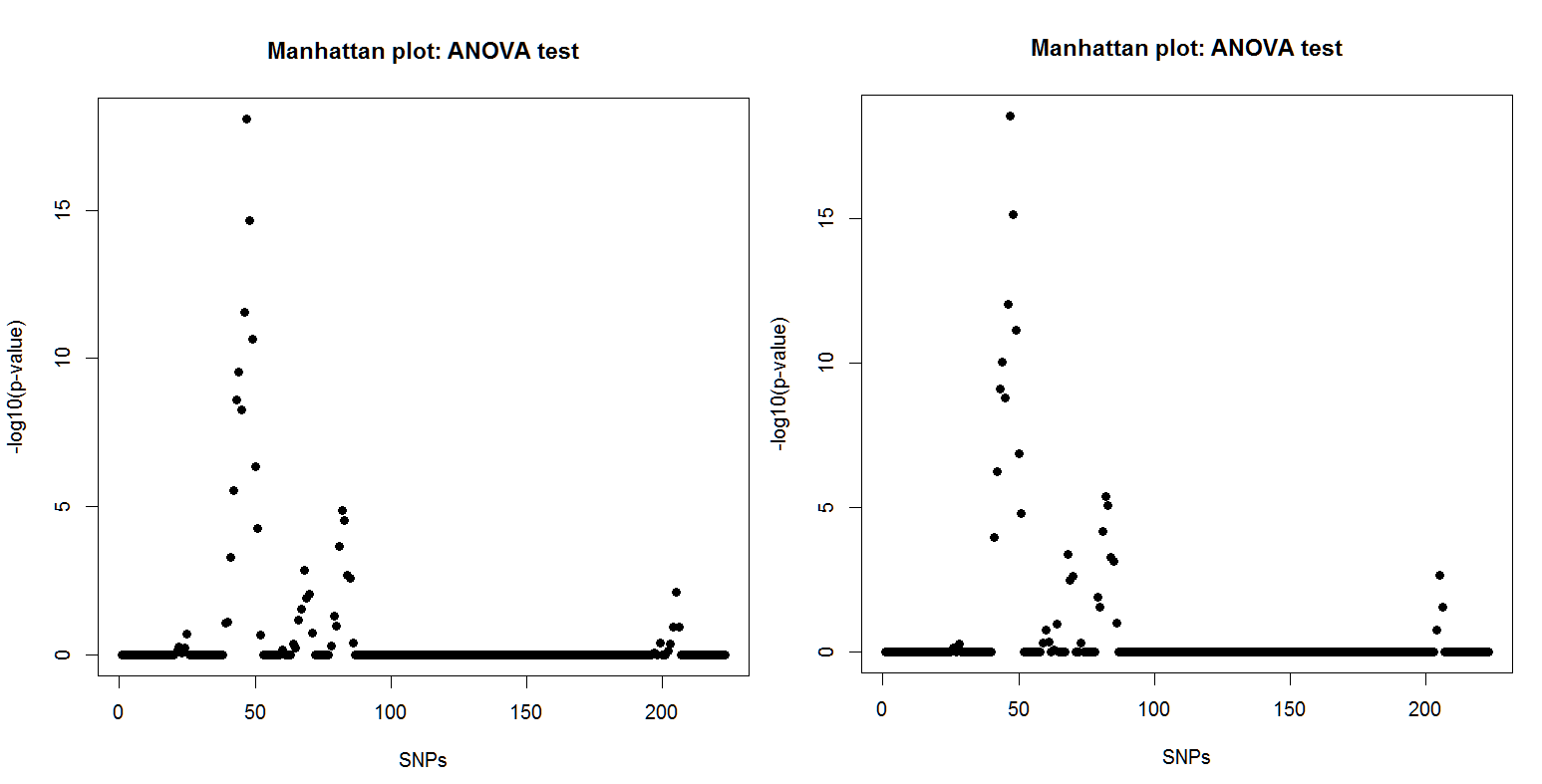}%
\caption{The Manhattan plots for the Steptoe x Morex data set (the heading
date trait) using standard method (left) and the FAPI (right)}%
\label{fig:Manh_plot_56}%
\end{center}
\end{figure}

The third dataset obtained from the Steptoe x Morex cross. First, we analyze
lines with respect to the heading date trait. According to the standard
ANOVA\ test, the 47-th SNP has the smallest p-value which is $8.5\times
10^{-19}$. Other significant SNPs have numbers 68, 82, 205. However, they have
larger p-values, namely, $1.48\times10^{-3}$, $1.37\times10^{-5}$,
$8.12\times10^{-3}$. The Manhattan plot generated from the obtained p-values
is shown in Fig. \ref{fig:Manh_plot_56} (the left plot). By using the FAPI, we
get quite the same results. The Manhattan plot generated from the p-values
obtained by means of the FAPI is shown in Fig. \ref{fig:Manh_plot_56} (the
right plot). Moreover, we obtain the strong interactions of SNPs $47\times82$
(p-value is $1.7\times10^{-26}$), $47\times205$ (p-value is $7.66\times
10^{-24}$), $47\times68$ (p-value is $2.07\times10^{-21}$), $47\times165$
(p-value is $4.65\times10^{-13}$), $47\times102$ (p-value is $1.07\times
10^{-12}$), $47\times134$ (p-value is $2.26\times10^{-12}$).

The standard analysis with respect to the grain yield trait gives the
following significant SNPs and their p-values in parentheses: 82
($2.69\times10^{-11}$), 20 ($1.03\times10^{-2}$), 68 ($3.02\times10^{-2}$),
129 ($4.04\times10^{-1}$). The FAPI provides the same significant SNPs.
Additionally, we get the following interacting SNPs: $82\times112$
($1.8\times10^{-11}$), $82\times151$ ($5.56\times10^{-8}$), $82\times135$
($5.12\times10^{-6}$), $82\times195$ ($1.22\times10^{-5}$).

\begin{figure}
[ptb]
\begin{center}
\includegraphics[
height=2.3633in,
width=4.6167in
]%
{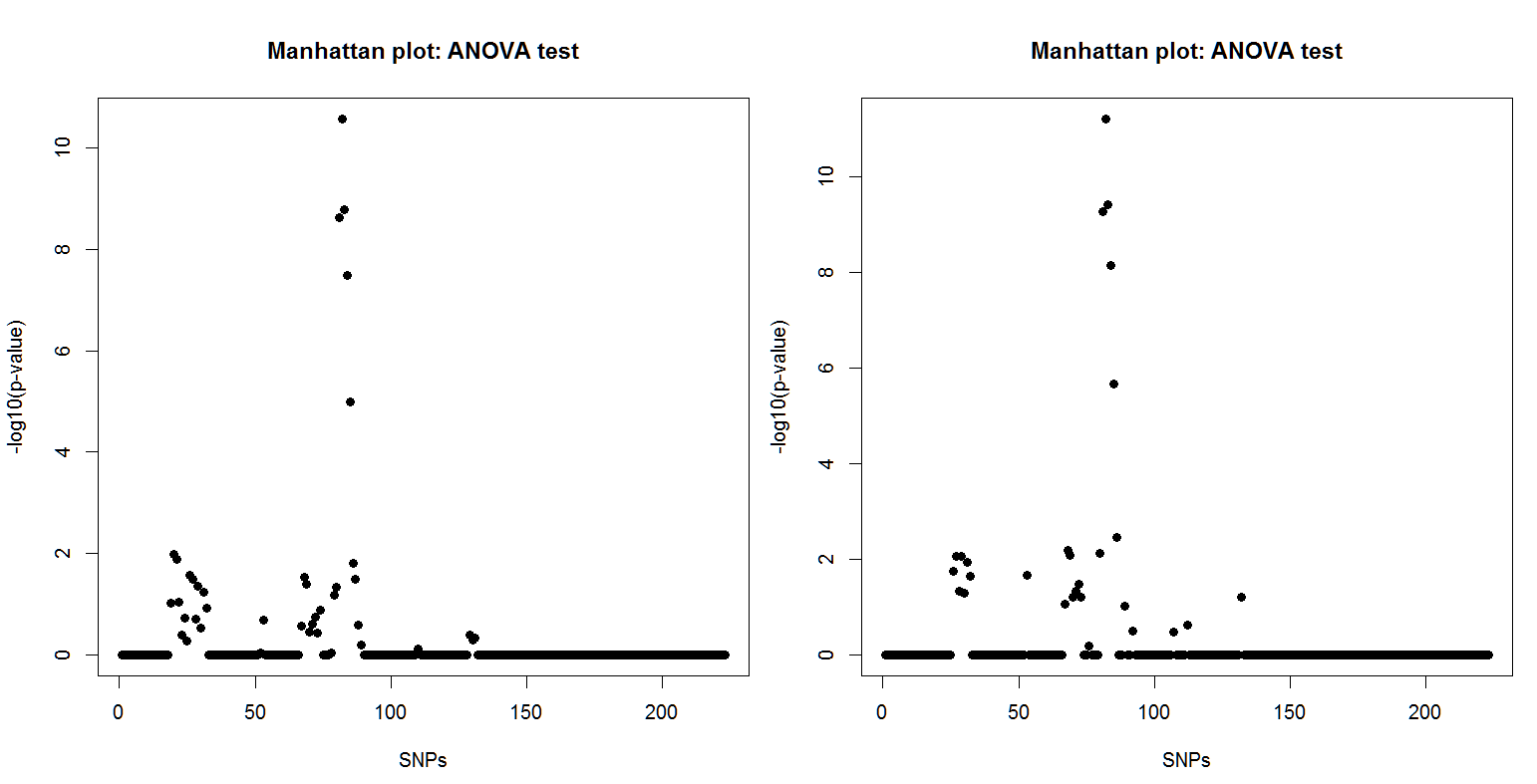}%
\caption{The Manhattan plots for the Steptoe x Morex data set (the grain yield
trait) using standard method (left) and the FAPI (right)}%
\label{fig:Manh_plot_78}%
\end{center}
\end{figure}

The corresponding Manhattan plots generated from the p-values for the grain
yield trait are shown in Fig \ref{fig:Manh_plot_78}.%

%

%

%

\section{Discussion of improving the algorithm}

Let us point out shortcomings of the FAPI and discuss possible ways to
overcome them and to improve the algorithm.

First, numerous experiments with real data illustrated that the FAPI selects
groups of adjacent strongly correlated SNPs in the same chromosomal region
which are not inherited randomly. This effect is similar to those taking a
place in the ridge regression algorithm which tends to select all of the
correlated SNPs and make their importance coefficients to be equal. In
contrast to the ridge regression, the Lasso method tends to select only one
SNP from the group of correlated ones. Therefore, the problem of correlated
SNPs can be solved by using a two-step procedure. The first step is based on
the FAPI. Results of this step is a small set of important SNPs. The second
step uses the Lasso method or its modification, for example, the adaptive
Lasso, in order to remove the correlated SNPs from the available small set.
Moreover, we can use a modification of the Lasso which takes into account the
epistatic effect because the number of possible pairs of SNPs after the first
step is rather small.

Another way to treat with the correlated SNPs is to use the standard tools for
testing the association between single SNPs and a continuous phenotype,
including for example, one-way ANOVA. In order to identify two-locus epistatic
effect or interacting SNP-pairs that have strong association with the
phenotype, an algorithm for the two-locus ANOVA test can be used. There are
many approximated methods for reducing the computational burden. They are
reviewed in detail for a case-control study when the phenotype can be
represented as a binary variable with 0 representing controls and 1
representing cases as well as for the quantitative trait locus analysis when
the phenotype is quantitative is provided by Zhang et al.
\cite{Zhang-Huang-Zhang-Wang-2012}. Most methods are reduced to two steps. The
first step is reduction of a sets of SNPs in order to apply standard
statistical procedures to this reduced set of SNPs. The standard statistical
procedures make up the second step. The reduction of the set of correlated
SNPs can be successfully implemented by means of the FAPI as the first step.
As a result, we get a small subset of important SNPs which can be processed by
statistical tests, for instance, ANOVA test, in order to remove the correlated
SNPs located on the same chromosome.

We point out another shortcoming which has been observed in numerical
experiments. Since the number of SNPs is much larger than the number of
individuals, then we observe only a very small number of vectors
$\mathbf{x}_{i}$ among all possible vectors. This implies that contributions
of some important SNPs in a pair of vectors of alleles $(\mathbf{x}%
_{i},\mathbf{x}_{j})$ may be hidden when there are many transitions in this
pair, for example, from $0$ to $1$ and from $1$ to $0$. In this case, the
distance between vectors is large, and this pair does not get to a set of $N$
\textquotedblleft best\textquotedblright\ pairs with the largest ratios
$r(i,j)$. One of the ways to overcome the difficulty is to apply the
combination of the bagging method \cite{Breiman-1996} and the random subspace
method \cite{Ho-1998}. The FAPI can be improved by using a combination of the
bagging method for individuals and the random subspace method for SNPs. The
random sampling of individuals in the proposed method allows us to
\textquotedblleft smooth\textquotedblright\ some outliers of the phenotype
caused by random factors. By means of the random sampling of SNPs, we try to
reduce the effect of SNPs which mask the effect of subsets of important SNPs.

\section{Conclusion}

In this paper, a very fast and simple algorithm for GWAS, including SNP
interaction detection, has been presented. In spite of its simplicity, the
FAPI can be applied to various GWAS problems and cases from the analysis of
binary genotype matrices to the microarray gene expression data analysis.
Moreover, the algorithm can be simply extended, for example, on the bagging method.

At the same time, it is important to note that the algorithm should be used
jointly with another algorithm, for example, with the ANOVA tests to identify
the association between a single marker or interacting SNP-pairs and a
continuous outcome. At that, the second stage uses a set of significant SNPs
which is obtained at the first stage by means of the FAPI.

The results of numerical experiments and the logic underlying the FAPI have
demonstrated that it outperforms the standard algorithms from the
computational point of view for many real data sets. Moreover, it takes into
account the epistatic effect or the SNP-SNP interaction. We have analyzed DH
populations of barley for purposes of numerical experiments. The experiments
have illustrated the FAPI efficiency. The obtained sets of significant SNPs
have coincided with similar sets obtained by means of standard algorithms.
Moreover, we could see that SNP-SNP interactions detected by means of the FAPI
were successfully validated by performing the two-locus ANOVA test. However,
we have investigated only rather small data sets and only a simplest
implementation of the FAPI. It has been done because we aimed to compare
results of the FAPI with the well-known standard technique. We aimed to get
added evidence that the algorithm copes with tasks of the GWAS. It should be
noted that a lot of experiments have to be performed in order to evaluate how
the FAPI handles various types of data set, large data sets, how its
modifications and extensions outperform the available algorithms. These
questions are directions for further research.

\bibliographystyle{plain}
\bibliography{Barley,Boosting,Classif_bib,Cluster_bib,Epistasis,FeatureSelection,GeneSelection,Lasso,MYBIB,MYUSE,Regres_learn}

\end{document}